\documentclass[pre,aps,preprint,amssymb]{revtex4}

\pdfoutput=1

\usepackage{amsmath}
\usepackage{tabularx}
\usepackage{pifont}
\usepackage{makeidx}
\usepackage{epsfig}
\usepackage{dcolumn}

\begin{document}

\title{Does the second law of thermodynamics really hold good
       without exception?}

\date{\today}
\author{Hans R. Moser \\
{\small Physics Institute, University of Z\"urich,}
{\small Winterthurerstrasse 190, CH-8057 Z\"urich, Switzerland} \\
{\small E-mail: moser@physik.uzh.ch}}

\begin{abstract}
A major part of the many thermally driven processes in our natural
environment as well as in engineering solutions of Carnot-type machinery
is based on the second law of thermodynamics (or principle of entropy
increase). An interesting link between macroscopically observable quantities
of an ensemble (state variables) and the thermal velocity of its individual
constituents such as molecules in a liquid is provided by the Brownian
motion of suspended larger particles. We postulate a "frustrated
Brownian motion" that occurs if these particles get partially trapped in an
environment of suitable geometrical conditions. This dissipates a small
fraction of the kinetic energy attended with the Brownian motion and
deposits it inside the trap, and so this constitutes a mechanism that by
itself transfers thermal energy from cold to warm. We note that this is just
a marginally admitted, slowly evolving effect driven by a thermodynamic
quasi-equilibrium, thus being of limited efficiency in terms of energy
density attainable per unit of time. However, a simple experiment suggests
that this process indeed is allowed to take place, and the envisaged
applications then are straightforward.


\vspace{1cm}


\end{abstract}

\maketitle

\section{Introduction}
The question whether it is possible to satisfy our energy needs without
to burn down some kind of fuel has attracted scientists over many centuries.
Apart from the use of other natural resources (mainly sun, water power,
and wind), the proposed ideas range from absolutely bizarre up to
ingenious. While it appears not too hard to accept conservation of total
energy including heat according to the first law of thermodynamics, the
second law certainly deserves a closer look at its practical implications.
We may ask where it poses an ultimate limit concerning energy exploration
in a technical or economical sense. Why is it impossible to cross a lake
with a boat just by cooling down the water in the lake? The second law
states that we need two reservoirs of heat at different temperatures, and
it has never been observed that we get by with just one of them, i.e., the
lake in our example.

Our above example is almost archetypal for lots of practical and
sometimes also theoretical questions that relate to the second law of
thermodynamics. We face a situation where the energy under consideration
is all present, but the second law prohibits access to it. In this
contribution we claim that there is a way out of this difficulty, i.e., our
approach spontaneously creates a temperature difference that may be used
in a traditional way. However, we also anticipate that our experimental
setup does not work too efficiently. Nevertheless, besides
aspects of principle, a numerical estimation clearly offers
a chance for an economically relevant implementation of a future power
supply, e.g., of an entire power plant. Further, to gain energy from the
environment undoubtedly would be beneficial to the climate. Including
the exhaust heat of a Carnot-like thermal engine that makes use of the
achieved temperature difference, we think of an energy economy that
(theoretically) is just neutral with all respects.

Concerning entropy, we refer to some of the forthcoming sections below. We
think it is hard to discuss this matter in full generality, while the
restriction to our specific experimental arrangement renders the pertinent
questions much easier. But we may already address the crucial issue, namely
the meaning of a local entropy sink with regard to the situation in
the environment.

The main goal of this work is to reconsider certain aspects of Brownian
motion within the particular context of the second law of thermodynamics.
On this score, we scrutinize some actually well-known peculiarities of
the Brownian motion in a less common framework of methods. The textbooks
\cite{kittel,blundell,roy} provide general thermodynamics, where \cite{roy}
also gives a little survey of technically relevant thermal engines.
Concerning a more specific focus on Brownian motion, we recommend
Refs.~\cite{nelson,kubo,marshall}. In a slightly enlarged context,
we also may look at particle motion in a homogeneous
liquid \cite{powles}, i.e., in the absence of larger
immersed Brownian particles. It is important to recognize
that the Brownian particles contribute to a thermodynamic (nearby) equilibrium
situation that comprises the host medium where they are accommodated.
Thus, the Brownian motion should not be considered to be "externally" powered
by the thermal energy of, say, a liquid water reservoir at some temperature.
Instead, the driving force emerges from the local and momentaneous deviation
from the statistical average of certain relevant properties the water molecules
may adopt, their momenta being of particular importance. Brownian motion
requires no external Newtonian driving force other than the one supplied
by the randomly occurring asymmetry in the local neighborhood of a particle.

Then, by chance, such a particle on its path may enter a suitably
prepared trap (see below). If many particles therein slow down, they accumulate
thermal energy at this place, since a small fraction of their kinetic energy
gets transferred to the local environment. Although the Brownian particles
are in thermodynamic equilibrium (actually it is a quasi-equilibrium, see below)
with the ones of the host medium, the trap may well react upon the different
size of particles. Our experiments suggest that this causes the Brownian particles
to slightly increase their tiny deviation from equilibrium.

Does this imply that we even might achieve a cyclically (perpetually) operating
thermal engine? The indispensable prerequisite to such a machine
is its ability to remove the particles from the trap in a way
that consumes less energy than they previously have supplied. We postpone
a further discussion of this matter until the experiment is described and
the measurements are explained within the scope of our suggested mechanism.
But we emphasize that already a partial realization, i.e., a partial
compensation of the energy needed to empty the trap, may well be greatly relevant
to technical applications. Admittedly, our performed experiment does not yet
comprise removal and reinjection of the Brownian particles. At present,
it covers only the single-step mode where the particles spontaneously transfer
energy along a temperature gradient in the "uphill" direction.

In a way, this resembles an everyday situation all of us possibly have
experienced now and then. The stormy weather in autumn whirls around the leaves
from the trees, and we consider the situation where a litter basket is present.
Initially, the walls of the basket prevent the leaves from entering, and so the
coverage (leaves per area) inside is lower than outside. But then,
under suitable overall conditions, there is even a crossover where the density
inside starts to exceed the one in the surroundings, since the walls now
capture the leaves inside the basket. Thus, we face a situation that
more or less pictures the one described above: within the basket (that serves
as our trap) numerous leaves (Brownian particles) are present. We recognize that
these leaves basically are at rest, since the litter basket limits their
available volume. Moreover, to far extent the filled basket also keeps off the
wind, i.e., there is almost no driving force anymore that might influence their
state in the sense of position and momentum.

\section{Theory}
\subsection{ Brownian motion and diffusion}
Although Brownian motion is a well-investigated subject matter known
for a long time, we shall briefly revisit some issues that will be important
below. The situation where a Brownian particle undergoes a (single)
collision with a neighboring molecule of the host medium may be described by
the Langevin equation
\begin{equation}
m\ddot{{\bf x}}(t) = -\beta\dot{{\bf x}}(t)+{\bf F}_c(t)+{\bf F}_{ext}.
\end{equation}
Here $-\beta\dot{{\bf x}}(t)$ denotes the usual friction term, ${\bf F}_c(t)$ is
the random driving force due to the collision, and ${\bf F}_{ext}$ means
a possible external force such as gravitation. The latter is negligible in our
situation of small particles (see below) that do not sediment to
the ground.

We may now perform the sum over many of these collisions, i.e., we consider
a longer period of the rapidly varying random force ${\bf F}_c(t)$ above.
This can be shown to be equivalent to the more
convenient sum over many (identical) particles. Further, we make use
of isotropy that removes the random term from the achieved sum, and we assume
also the equipartition theorem to hold true in this case. Then, after some
steps of rearrangement we arrive at an equation for the spatial mean value
of the sqared distances the particles have departed from their starting point,
namely
\begin{equation}
\frac{d^2}{dt^2} \langle {\bf x}(t)^2 \rangle_x
+ \frac{\beta}{m} \frac{d}{dt} \langle {\bf x}(t)^2 \rangle_x
= \frac{2kT}{m}.
\end{equation}
In our particular situation of Brownian motion, equipartition
relates the mean kinetic energy of a particle to its thermal energy. With
regard to the three translatory degrees of freedom, for $N$ (in ideal
circumstances spherical) particles this reads
\begin{equation}
\sum_{i=1}^{N} m \dot{{\bf x}}_i(t)^2 = 3NkT,
\end{equation}
where $k$ and $T$ are the Boltzmann constant and temperature, respectively.
Clearly, $N$ in Eq.~(3) must be large, such that the sum does not fluctuate
anymore. For small $N$ or even for a single particle, we have to replace
$\dot{{\bf x}}(t)^2$ by its time average $\langle \dot{{\bf x}}(t)^2 \rangle_t$.

The stationary solution of Eq.~(2) is recognized to grow linearly in time. In
the most common notation it may be written as
\begin{equation}
\langle {\bf x}(t)^2 \rangle_x = 2Dt,
\end{equation}
with the diffusion constant
\begin{equation}
D = \frac{kT}{\beta} = \frac{kT}{6\pi\eta r},
\end{equation}
where the second part of (5) applies to the important special case of
spherical particles in a liquid. Moreover, the viscosity $\eta$
exhibits quite a strong dependence on temperature, even in water.

We think it deserves attention that, even in our context of Brownian motion,
the equipartition principle nicely stands an experimental
confirmation of its validity. Generally it is the Avogadro number
that gets determined in a rather indirect manner, and
the result turns out to be correct within an almost amazing precision of
roughly one percent. In our view this is not really evident. Over a large
range of size and mass of the Brownian particles, they carry a mean
translatory kinetic energy of $(3/2)kT$ each, just as the other particles
(e.g., water molecules) of the host medium do.

\begin{figure}

\vspace{-3.2cm}

\hspace{-3.0cm}\includegraphics[scale=0.92, angle=0]{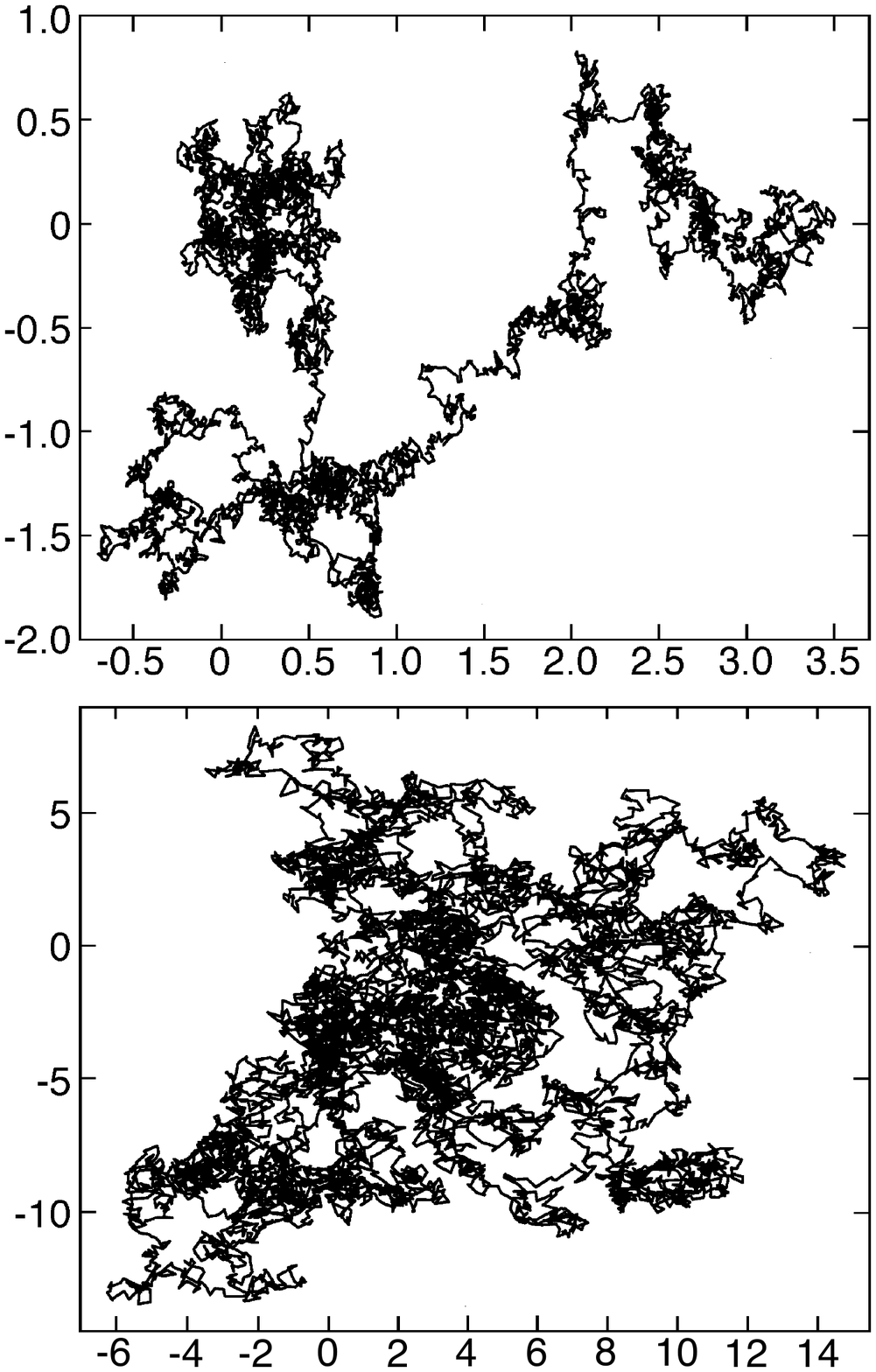}

\vspace{-4.2cm}

\caption{Top: Two-dimensional Brownian motion based on
$2 \times 10^6$ points of a trajectory ${\bf x}(t)$ that solves
Eq.~(1), $10^4$ out of them being plotted. Bottom: This time
we use $2 \times 10^8$ points, again $10^4$ of them are displayed,
and so the resolution is reduced by a factor of $100$. If we
compare the two panels, we note self-similar properties
(see text) as well as spatial extents according to Eq.~(4).
\label{fig1}}
\end{figure}

Now we may focus on another important property of the Brownian motion,
namely its fractal scaling behavior. This emerges from an inspection of
the particle trajectories at different scales of length and time. Equation (4)
suggests that, provided fractal behavior is present, to rescale spatial
extents by some factor of $A$ and the time by $A^2$ should preserve the
global aspects of a Brownian trajectory (but not its local details).
Figure~1 presents such trajectories, namely numerical solutions ${\bf x}(t)$
of Eq.~(1), which already exhibits the major issues of this circumstance.
We simulate the driving force ${\bf F}_c(t)$ in (1) by
truly decorrelated random numbers, and so we look at the
special situation of a Wiener process. The bottom
panel shows a two-dimensional Brownian motion based on $2 \times 10^8$
points or time-steps, where the resolution is such that $10^4$ points are
plotted. The upper part of the figure pictures the very same motion with
only $2 \times 10^6$ points, and again $10^4$ out of them being used to
display the curve. In other words, the upper panel shows the first percent
of the bottom trajectory, but the resolution above is $100$ times better.
We observe the roughly $10$ times larger linear spatial extents in the
bottom panel, while it covers a period of time that is as much as $100$
times longer. This applies also to a situation in one or in three dimensions,
since a move into, say, the direction perpendicular to Fig.~1 does not
depend on the motion within the plotted two dimensions.

At first glance, the agreement appears to be rather poor: the upper motion
exhibits noticeable excursions, whereas the bottom situation looks more
compact. However, this impression is not really significant. The two
trajectories start at (0/0) in their plotted frames, and so we recognize
that the upper curve initially represents quite a local blur.
Moreover, due to the different scales of length such excursions seemingly
are less pronounced in the bottom panel. Altogether, Fig.~1 is
meant to simulate the quite well defined ordinary situation
outside the particle trap, i.e., the reference case in the
absence of an obstacle to the Brownian motion. We anticipate that,
due to insurmountable computational hurdles, most of the differences
occurring inside the trap must be quantified on an experimental basis.

Solution (4) tells us also that we deal with phenomena in a diffusion context.
The peculiarities of Brownian motion theory appear just from a detailed
inspection of a single particle motion, rather than of a particle density
$n({\bf x},t)$. This way we also may recognize the Brownian particles to be
in a quasi-equilibrium with the ones of the host medium. Diffusion processes
are time-dependent evolutions but actually they are rather slow, and so this
just insignificantly affects the equipartition energy value of $(3/2)kT$.
Usually, diffusion processes are described by two major statements, namely
\begin{equation}
{\bf j}_n({\bf x},t) = -D\nabla n({\bf x},t)
\end{equation}
and, assuming there is no particle source other than the initial one,
\begin{equation}
\frac{\partial n({\bf x},t)}{\partial t} = D\Delta n({\bf x},t).
\end{equation}
The vector ${\bf j}_n({\bf x},t)$ in (6) is the particle current
density, and $\Delta = \nabla^2$ denotes the Laplacian operator.

Sometimes Eq.~(7) is generalized in the sense that all the
processes that formally adopt this type of dependence on space and
time are said to be "diffusion-like". Probably heat conduction is
the most common example, but there is a variety of phenomena where
a gradient in some quantity tends to even out in a diffusion manner.

Actually, the modified or retarded diffusion mechanism inside the
particle trap (see experimental section below) may well be regarded as
a further example of generalized diffusion. Concerning the
trajectory ${\bf x}(t)$ of an individual particle, this is what we call
"frustrated Brownian motion". The numerous interaction events of such
a particle with the internal surfaces of the trap involve dissipative
(inelastic) contributions, which is significantly beyond
what we might tackle in terms of dynamical systems modeling. These
interactions hardly may be incorporated into an equation of type (1),
and so we cannot offer a modified plot in the spirit of Fig.~1. In all
probability such an attempt would not be able to care for the exact
geometrical situation of an individual interaction process, thus providing
us just with a summary effect (based on assumptions) of the trap.
We rather might try to generalize Eq.~(7) up to the extent where it
reproduces the measured overall effects (for all the Brownian particles).
But this necessarily means that we parametrize our diffusion model, and
particularly the appearance of a rather faint effect likely may emerge
from a suitable parameter choice.

A possibility for such generalization steps may be the Fokker-Planck
equation~\cite{kubo} that, in addition to Eq.~(7), comprises a drift
term. But this will by far not suffice to simulate the experimental
situation inside the cavities of the trap, and so it definitely cannot
bypass the need for free parameters. Therefore, we think an experiment
should answer the open questions.

Nevertheless, our idea of dissipated translatory particle energy
inside the trap is largely motivated by diffusion (or Brownian motion)
theory itself. We face a particle density gradient that
drives the diffusion mechanism and that inherently represents
a quasi-equilibrium. The trap then may cause the Brownian
particles to depart somewhat stronger from thermodynamic equilibrium energy
(particles plus host medium), namely from the equipartition value of
$(3/2)kT$ per particle. After all, we think this energy difference between
quasi-equilibrium (usual diffusion process outside the trap) and actual
situation based on a slightly more pronounced deviation from equilibrium
(inside) produces our measured effect, i.e., the temperature gradient
under consideration.

\subsection{Fluctuations in general}
Since in a diffusion context we deal with a quasi-equilibrium, we
should be aware of an important link between equilibrium and
non-equilibrium statistics, namely the celebrated fluctuation-dissipation
theorem. Essentially it states a relation between thermal fluctuations and
the response of a system to an external disturbance. For a Brownian
particle the theorem says that the dissipative frictional drag the particle
undergoes upon an externally powered dislocation has the same physical
origin as the Brownian fluctuations themselves. This is also the basic
assumption to (strictly) derive the first part of Eq.~(5). A thorough
discussion may be found in \cite{blundell,kubo}, but here we
merely state the particularly simple and experimentally well-investigated version
\begin{equation}
\tilde{P}(\omega) = \frac{2kT}{\omega} \rm{Im} \tilde{\chi}(\omega).
\end{equation}
$\tilde{P}(\omega)$ denotes the power spectral density, namely the Fourier
transform of $x(t)^2$ that is a squared "system output" with almost
no restriction to its physical meaning. Since the squares imply
a nonzero time average, Eq.~(8) requires the situation of a stationary
random process: the Wiener-Khinchin theorem then permits the
Fourier transform of $x(t)^2$ in terms of a transformed
autocorrelation function. Further, $\tilde{\chi}(\omega)$ is the
Fourier transformed susceptibility or linear response function $\chi(t)$.
Including $\Theta(t-\tau)$ that is the Heaviside function, $\chi(t)$ most
conveniently may be introduced by the statement
\begin{equation}
\langle x_{resp}(t) \rangle_t = \int^\infty_{-\infty} \Theta(t-\tau)
\chi(t-\tau) f(\tau) d\tau.
\end{equation}
Here $f(t)$ means a generalized scalar external force, and
$\lambda H^1(x,t) = x_{resp}(t)f(t)$ denotes the corresponding time-dependent
perturbation in the overall Hamiltonian function
$H(x,y, ... ,t) = H^0(x,y, ...) + \lambda H^1(x,t)$. Clearly, $f(t)$ must be
organized in a way that $H^1(x,t)$ adopts the dimension of an energy.
To first order in $\lambda$, we recover the response $x_{resp}(t)$ to $f(t)$
that is linear in $f$ (just as, e.g., in a linear spring), in accordance
with Eq.~(9). The Heaviside function assures causality,
i.e., the effect of the force $f(t)$ that acts at some point of
time cannot appear prior to this very moment. Generally $x_{resp}(t)$ may
depend on all the past values $f(t)$ for $t < t_0$ up to some time $t = t_0$
of interest, say, up to present. This yields a progressing time-average
$\langle x_{resp}(t) \rangle_t$ that may strongly depend on $t_0$ and
adds to the (time-independent) mean value $\langle x_0(t) \rangle_t$ of the
unperturbed signal. Thus, we arrive at the observable "up to present
average" $\langle x(t) \rangle_t$ of the total system's output. Then, the
imaginary part of $\tilde{\chi}(\omega)$ in statement (8) describes the
dissipated energy caused by $f(t)$, while $\tilde{P}(\omega)$ characterizes
and quantifies the fluctuations.

It has been known for a long time (actually since the pioneering
investigations of L. Boltzmann) that on small scales in space and time
the second law of thermodynamics fails. See Ref.~\cite{wang} for
a recent experiment where the deviations from equilibrium thermodynamics
become quantified. In a way, such approaches always relate to the
fluctuation-dissipation theorem (however, the authors in \cite{wang}
prefer a version of their own that facilitates a comparison with experiment). For
a theoretical discussion of finite size effects we recommend \cite{sekimoto}.
There, some pertinent quantities and processes that usually appear in
a context of macroscopic equilibrium thermodynamics are reconsidered for
ensembles with limited heat capacities, where the thermal fluctuations
cause additional implications. In such cases, to achieve the correct
limit for infinite size and/or time is of particular importance.

The short-term production or consumption of entropy prior to equilibrium
necessarily involves a time-dependent entropy definition, we refer to an
instructive treatise on this matter \cite{mackey}. Most commonly, the
time-dependence gets introduced by an intuitive generalization of the
Gibbs entropy stated in terms of an evolving $N$-particle probability density
$W_N$, namely
\begin{equation}
S(t) = -k\int W_N({\bf r}_i,{\bf p}_i,t) \ln W_N({\bf r}_i,{\bf p}_i,t)
d^3r_i d^3p_i.
\end{equation}
The index $i$ in Eq.~(10) is meant to denote the totality of particle labels,
and we assume $N$ to be the number of all the particles (Brownian ones plus
host medium) within some experimentally relevant piece of volume.
Further, definition (10) ignores degrees of freedom other than translatory
ones and, hence, the integration is thought to cover a $6N$-dimensional
phase space. The time-dependence in $S(t)$ is not primarily supposed to
characterize fluctuations in small systems: it rather applies to macroscopic
ensembles that have not yet arrived at their thermodynamic equilibrium.
Then, Ref.~\cite{mackey} exemplifies that a system's entropy
in an off-equilibrium situation does not always grow monotonically,
thus may exhibit temporary minima, and so we may ask whether our
suggested violation of the second law is nothing but a momentaneous
entropy loss that later on gets compensated.

However, we urge the reader to recognize the time-dependence in
a diffusion experiment (see below) to be notoriously slow and, on top of
that, we observe the outcome of a macroscopic thermal situation that
largely is spatially averaged. In simple terms, the deviations from
thermodynamic equilibrium are minute, and so they definitely cannot account
for too prominent time variations. Thus, a drop in the overall
system's entropy surely cannot be attributed to a temporary event due
to some kind of fluctuations. The concluding section below addresses
entropy accounting in our specific experimental setup in terms of
equilibrium thermodynamics. Admittedly, our proposed dissipation of kinetic
energy within the trap implies that the Brownian particles leave the
equilibrium with the liquid (although the experimental deviation
is really a minor fraction of the equilibrium energy). But this is
definitely not a fluctuation in the spirit of theorem (8), and in all
probability it is not a temporary entropy minimum (in $S(t)$ above) on the
way to a maximum that will persist. We cannot continuously take advantage of
an entropy minimum appearing significantly out of an equilibrium situation.
The experiment below in fact has the power to decide where the additional
thermal energy in the particle trap originates from, namely from the
surrounding liquid where the trap is located. Hence, at given particle
density distribution the measurable temperature gradient lowers the total
entropy content in the overall (closed) system, but this is just a scarcely
admitted marginal deviation from equilibrium. Similarly, the second law
in the version where it prohibits a perpetually operating
machine is only marginally supported by statistical thermodynamics.

Clearly, the dissipation process comes to an end after
some time, but this happens essentially because the filled
trap cannot accept further particles anymore. Then, expectedly, the
experiment exhibits a subsequent thermal compensation flow that equilibrates
the achieved temperature gradient (and thus produces entropy). Therefore,
this entropy contribution is just a feature of our present single-step
experiment, and the costs to empty the trap in a continuous manner will be
discussed later on in the experimental section. Very importantly, even during
the crucial phase of the experiment where the particle trap accumulates
heat at the cost of the surroundings, we deal with a diffusion-like
process attended with a quasi-equilibrium.

We already stated some of the arising difficulties on the way to
dynamical modeling of the measured temperature differences.
Actually we lack both a microscopically sound description of
the interaction events between Brownian particles and the trap,
and also their summary effect in terms of temperature and entropy.
We are aware of recent theoretical developments that trace back
a macroscopic system's entropy to the behavior of its individual
particles along their trajectories \cite{seifert}, which also
extends to numerical examples \cite{xiao}. These issues are intimately
interwoven with the fluctuation-dissipation theorem (8) and its
alternate versions, consult also the pertinent references quoted in
\cite{seifert,xiao}. Apparently the
statistical limit for many-particle systems correctly
yields the ensemble entropy $S(t)$ above, and so it
seems to be adequate to define a trajectory-dependent entropy $s(t)$. Such
a "single-particle entropy" is defined in quite an intuitive way
as $s(t) = -k \ln p({\bf x}(t),t)$. Here $p({\bf x}(t),t)$ means the
probability (per volume unit) to find the particle, e.g., at the
moment $t_0$ near ${\bf x}(t_0)$.

With regard to modeling and numerical simulation, in fact this meets our
key problem. We intend to understand a Brownian particle that (above all
within the trap) is in an off-equilibrium situation, although being in
contact with a heat reservoir. At present we recognize a growing
literature on fluctuation theorems generalized to
non-equilibrium situations, as an example see Ref.~\cite{chetrite}.
Here it is crucial to distinguish the global non-equilibrium in the
host medium (heat reservoir) from the one that refers to a single
particle with respect to its neighbors in an
ensemble, both issues matter in our experiment.
The time-evolution of the above $s(t)$ (and thus also
of the envisaged $S(t)$ for an experimentally relevant ensemble) should
consistently incorporate entropy consumption due to our
measured energy transfer from outside into the trap. However,
our present knowledge about "frustrated" Brownian trajectories
inside the trap is clearly insufficient in order to turn this new
type of information into a conclusive numerical simulation of our
measurements. From a theoretical as well as numerical viewpoint,
quasi-equilibria are inherently difficult to handle, since the
macroscopic phenomena or quantities we are after always represent small
disturbances on top of a strongly dominating background situation.

Besides diffusion, we would like to state an everyday example of
a thermodynamic quasi-equilibrium, namely the macroscopic air flow
in the atmosphere. Such weather events may well reach the strength of
a hurricane, but actually this is just a minor deviation from
thermodynamic equilibrium. Probably nobody is that small-minded to claim
an undefined or ill-defined temperature due to non-equilibrium.
But it is also true that the deviation from equilibrium
rapidly increases with the strength of a storm, because of the
velocity square in the kinetic energy of the air flow. We think it is
illustrative to compare the numerical values of such kinetic energies to
the thermal ones, also in a diffusion context. The atmosphere provides
excellent opportunities to study self-ordering or pattern formation
phenomena. There exist also many laboratory systems within this context
and, as far as we can see it, all of them deal with a quasi-equilibrium
as an indispensable prerequisite.

In a supplementary remark we point to the quantum mechanical analogue to
diffusion processes that is factual and certainly exceeds a merely formal
correspondence to Eqs.~(6) and (7). Apart from some restrictions that
we shall specify below, the time-dependent Schr\"odinger equation
\begin{equation}
i\hbar \frac{\partial \psi({\bf x},t)}{\partial t}
= -\frac{\hbar^2}{2m}\Delta\psi({\bf x},t)
\end{equation}
may be recognized to be well within the scope of this matter. The right-hand
side in (11) presents the free particle Hamiltonian that acts
on the particle's wave function $\psi({\bf x},t)$. Definitely
$\psi({\bf x},t)$ is not yet a density in the spirit of its modulus
square, and there is also a complex prefactor in Eq.~(11). Nevertheless,
a decaying quantum mechanical wave packet has much in common with a diffusion
process. In the end we experience the spreading of a particle density or, in
the single particle case, of a probability density. It is essential that
a considered particle is free: if we admit a Hamiltonian supplemented with
a potential $V({\bf x})$ that captures, say, an atomic electron within some
local environment, the diffusion analogy breaks down and then we face
entirely different quantum phenomena.

Actually there are various interesting links between
Schr\"odinger wave mechanics, classical Newtonian dynamics,
and also statistical thermodynamics, for a recent review
see \cite{gorin}, and so in our view it should be possible
to transfer some of the here presented issues into a quantum
mechanical context. In many respects this also relates to the so-called
quantum Brownian motion \cite{hanggi}. These viewpoints,
however, are not really the subject matter of the present article, but we
would like to address such questions on a more speculative basis.

\section{Experiment}
\subsection{Intention and setup}
The purpose of our performed experiment is twofold. First, we would like
to confirm our claims and to prove the feasibility of our suggestion
that a device based on the described physical effects really is able
to operate. Secondly, modeling and simulation of the used type of trap
for the Brownian particles (see below) in our view is close to impossible
without introduction of free parameters. An experiment closes this gap and
may guide us to firm conclusions.

\begin{figure}

\vspace{-7.2cm}

\includegraphics[scale=0.75, angle=0]{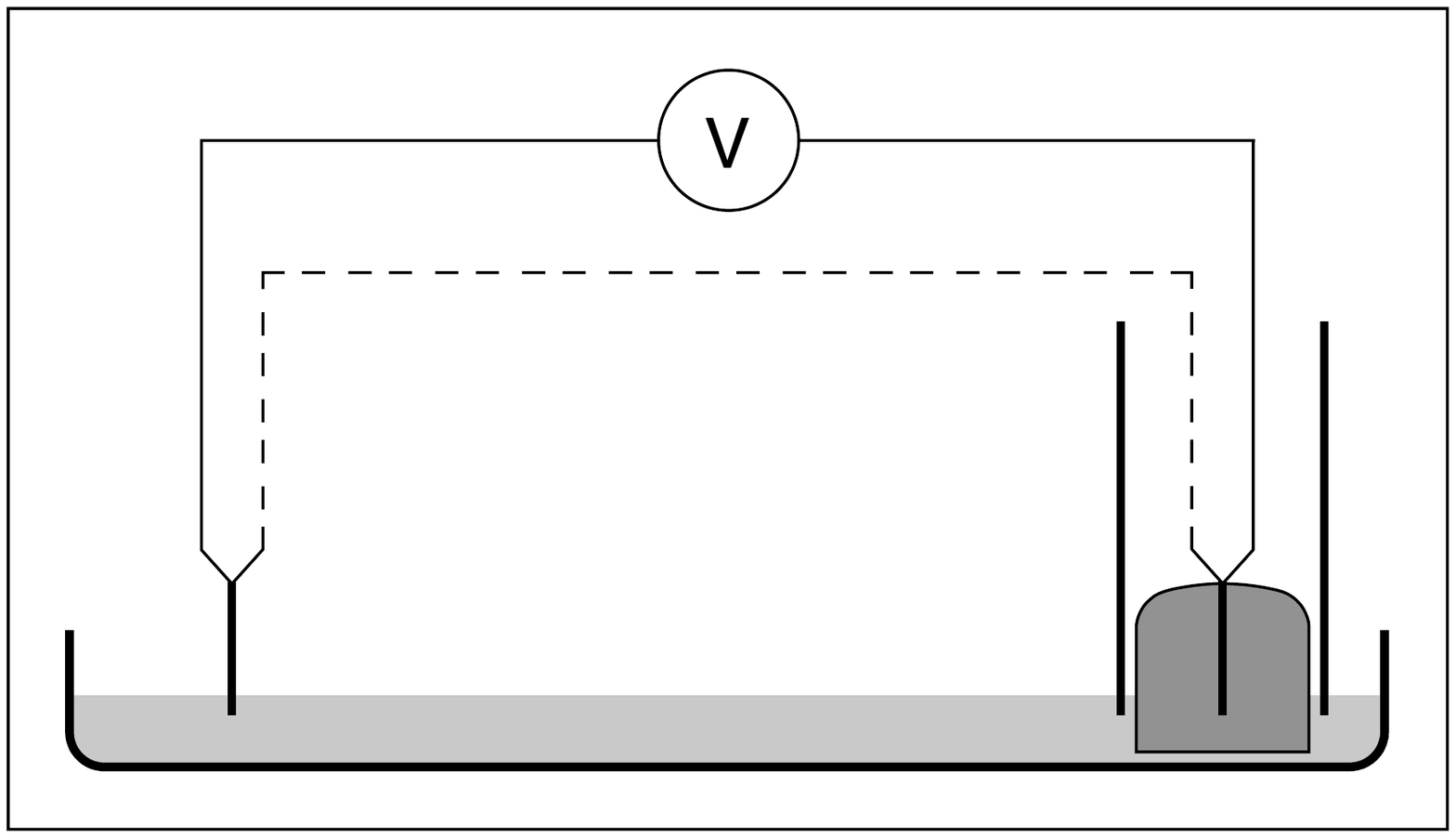}

\vspace{-8.0cm}

\caption{Experimental setup located in a Petri dish. To the right
there is our particle trap that merely features a plug of glass
wool inside a glass tube. Further, we point to the two sensors of
the thermocouple, where the one outside the trap serves as
a reference to the expected temperature increase due to dissipated
kinetic energy of the Brownian particles. The full and dashed
wires connected to the sensors refer to either alloy of the
thermocouple.
\label{fig2}}

\vspace{0.6cm}

\end{figure}

Figure~2 presents the entire setup. We use a Petri dish of 10~cm
diameter that is filled with 30~cm$^3$ water. Within some reasonable
limits such numerical values are not that critical in the sense that they
essentially affect the quantitative details of our plotted results. To the
right we place a glass tube of 1.5~cm inside diameter, and therein we put
a plug of glass wool as it frequently gets used for filtering purposes in
chemistry. The plug of glass fibers serves as a particle trap, and so
it certainly is one of our most important ingredients. Further,
there are the two sensor electrodes of a chromel-alumel thermocouple that
measure possibly occurring tiny temperature differences. These electrodes
have a length of 1~cm each and are soldered to the contact points
between the two different alloys of the thermocouple. The output voltage
(due to the Seebeck effect) amounts to approximately 40~$\mu$V/K with
quite a linear characteristic.

At some moment, roughly at the foremost point of the Petri dish
as it is drawn in Fig.~2 (closest to the observer, i.e., at
equal distance to the two sensors of the thermocouple), we supply
a few tenths of a gram potassium permanganate crystallites. Once these tiny
KMnO$_4$ crystals are dissolved in the host medium, they exhibit quite
a strong color close to the one of raspberries, a circumstance that will
prove extremely helpful (see below). The exact amount of supplied crystallites
actually is of minor importance, since we are always in a range where
some surplus KMnO$_4$ is left over in solid form after the experimentally
elapsed period of time. Then, the hydrated ions of the dissolved
potassium permanganate are our Brownian particles.

Since these particles represent two sorts of ions that, besides
their opposite charge, are unequally sized, we actually deal with
a modified version of "true" Brownian motion based on, e.g., colloidal
neutral objects with (preferably) a $\delta$-shaped size and mass
distribution. However, this circumstance will not affect the forthcoming
considerations. The hydrated ions are also subject to equipartition and,
on a large range of length scales, they diffuse in a very similar manner
compared to neutral particles. Figure~1 illustrates how they by
accident may end up in the particle trap. As indicated further above,
the idea of the experiment is to measure the accumulation of thermal energy
inside the trap, since therein we expect the Brownian particles to
dissipate a small part of their kinetic energy.

The color of the KMnO$_4$ solution should be seen as an essential part of
the experiment. It admits an observation of the two different time scales
within and outside the trap, respectively. Inside we experience
a propagation velocity in the order of 1~cm per hour, while elsewhere it
rather amounts to 1~cm per second. These velocities refer to the outermost
particle front that can be observed, and not to an average distance
$\langle {\bf x}(t)^2 \rangle_x^{1/2}$ in the bearing
of Eqs.~(2) and (4). Probably most other imaginable
particle types color too faintly in order to enable us to track how
they creep into the glass wool. Since potassium permanganate quite strongly
tends to oxidize combustible objects (and then the MnO$_4$ ions reduce to
manganese dioxide), we should pay attention that only sufficiently inert
materials (e.g., for the temperature sensors) get in touch with the liquid.
For example, a particle trap made of cotton wool is inadequate.

Further, the particle size deserves some special comments. The theory
section above explains that all the particles, regardless of their
size, have an average kinetic energy of $(3/2)kT$. We might consider
Indian ink with its size spectrum of the carbon particles between diameters
of some 0.1~$\mu$m (microns) up to more than 100~$\mu$m. In principle this
works just as well, but since these particles are that big we cannot get
sufficiently many of them into a given volume of water. This then limits the
attainable energy density within the particle trap. On top of that, the
Brownian motion becomes inconveniently slow. There is a wealth
of other ink types that split up into greatly different raw materials, but
generally their particles are also too large. Then, there are micelles
with a typical diameter range of, say, 10~nm (such as the ones in soapy
water) up to 100~nm, and so again this is somewhat too much. We may try
other ionic crystals but, apart from size effects, their thermal properties
upon dissociation and dilution in a solvent vary dramatically. Finally, our
choice of potassium permanganate proves alright, although there may be
many other possibilities we have not yet investigated. The
(hydrated) ionc particles are significantly larger than the
water molecules of the host medium, which is a prerequisite
to our simple type of particle trap. On the other hand, they are not
too large, in order to gather them in an adequately high density.
Thus we are left with a compromise that may well be further optimized.

From a technical viewpoint, shape, size, and density of the hand-made
plug of glass wool is the most critical matter in the experiment. Concerning
reproducibility, it must be clear that we cannot repeatedly manufacture such
a plug with the very same properties. For example, if the glass wool is
too tightly stuffed into the glass tube, the Brownian particles are too
strongly hindered from entering the trap. (The water, however, immediately
is there due to capillarity.) If the process of particle accumulation
works too slowly, there is enough time to cool down and hence to destroy
a measurable effect.

In fact it is the major problem in our experiment that the
outcome quite sensitively depends on the details of the plug.
This arises from the circumstance that we deal with several simultaneously
occurring heating and cooling effects. There is the endothermic solution
enthalpy (that comprises the exothermic hydration heat), the exothermic
enthalpy of dilution as the particles move on, the thermal energy due to
our expected effect inside the trap, and also thermal equilibration
as time goes by. Note that many textbooks on physical chemistry
incorporate the term "enthalpy of dilution" into a somewhat more general
concept, namely enthalpy of mixing. The asymmetry due to the plug of glass
wool in Fig.~2 provides an opportunity for all the ocurring thermal effects
to become manifest in the measured time-dependent temperature difference.
At first glance, the experiment looks quite simple, but then we face
a nontrivial balance between several thermal energies that strongly depends
on various geometrical conditions, particularly the ones inside the trap.

Conversely, to perform the experiment is really cheap, and so we may
repeat it many times. Reproducibility then refers just to the qualitative
shape of the measured curves (see below). The details, however, depend on
the particular properties of the plug used at a time. Although this
circumstance poses an experimental difficulty it still can be handled,
thus in our view it cannot depreciate or disqualify our statements
concerning the second law of thermodynamics. An obvious advantage of the
glass wool is the possibility to immediately modify certain properties
such as its density. In a future stage of the experiment, however,
we think the plug of glass wool probably gets replaced by some
well-defined type of micro- or nanostructures, and so the above mentioned
problems hopefully will drop.

\subsection{Measurements}
At this point, we would like to present our results that are summarized
in Fig.~3. All the curves are recorded in a way that the first 30 min show
the stability before supply of the KMnO$_4$ crystallites at $t = 0$. We
always plot the measured thermoelectric voltage $U$ between the two sensors of
the thermocouple versus time. The sign has been chosen such that an increase
in this electric potential difference means the particle trap gets warmer
with respect to the reference sensor.

Then, the conversion into a temperature difference is given by the further
above stated value of 40~$\mu$V/K due to the type of thermocouple, or 1~$\mu$V
on the ordinate axis means 0.025~K. Thus, our observed effects are in the
order of 0.1~K that already poses certain stability problems with regard to
temperature drifts as well as voltage measurement reliability. However,
the difference measurement cancels some of these difficulties. The recorded
points are spaced by mesh-intervals of 5~min and the resolution is
1~$\mu$V, and so the points cover a range of $\pm$0.5~$\mu$V around their
plotted values.

We might significantly improve the quality of the data by means
of a much higher sampling rate. A subsequent convolution with
an (e.g., Gaussian) window function of adequate width then would act
as a low-pass filter, as we performed it for rather noisy input
signals \cite{moser}. There we also performed a more advanced
filtering procedure based on a nearest neighbors search. This time,
however, we were not equipped on the same level of data
acquisition. The fluctuations are recognized to be largely
within $\pm$1~$\mu$V around the expectation values, which actually does not
conflict with the size of the discussed structures. Once we have experienced
the approximate width of the relevant features in the measured signal, we may
also consider to cut off the low-frequency noise. In Fig.~3, however, we
prefer to show the plain measurements, and we renounce a numerical procedure
in order to prettify the plotted curves.

\begin{figure}

\vspace{-4.0cm}

\hspace{-1.3cm}\includegraphics[scale=0.84, angle=0]{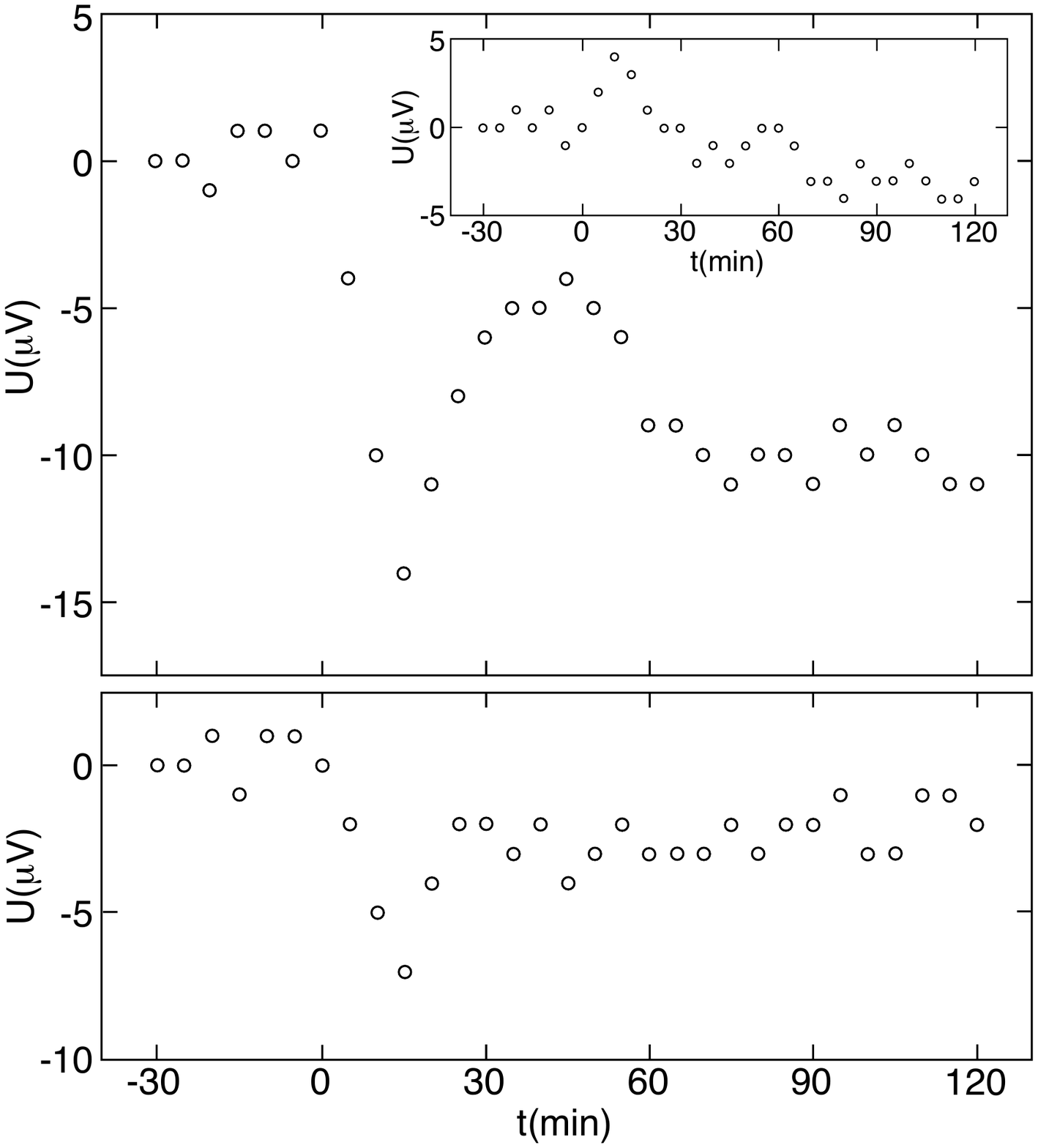}

\vspace{-5.0cm}

\caption{Top: Temperature difference between the two sensors
as a function of time, see text. The recorded thermovoltage
amounts to some 40~$\mu$V/K and the KMnO$_4$ crystallites have
been supplied at $t = 0$ on the abscissa. Besides the initial
trough (mainly due to dilution heat) we observe quite a pronounced
structure that quantifies the dissipated translatory energy of
the Brownian particles. Bottom: Reference plot based on a slightly
modified particle trap that avoids appearance of the effect. The
inset offers another type of comparison where one of the sensors
is kept at constant external reference temperature.
\label{fig3}}
\end{figure}

We also should be aware of a "pitfall" that arises from the difference
measurement. In plain terms, we cannot distinguish whether a sensor gets
heated or the other one cools down. On the other hand,
this works much more precisely than two independent temperature
measurements, since the small temperature effect within the
particle trap happens relatively to the environment elsewhere
in the Petri dish. The upper panel in Fig.~3 exhibits the effect we are after,
as we shall explain below in this section. The bottom curve is recorded under
the same conditions, except that the particle trap is slightly modified in
a way that the effect does not appear. Therefore, the bottom panel serves
as a reference to the top one. Additionally, the inset in Fig.~3 presents
a temperature measurement inside the trap relative to a constant external
reference temperature.

In the upper panel we recognize quite a conspicuous trough around
$t = 15$~min that mainly emerges from warming up the reference sensor
electrode. Initially dilution heat seems to dominate other contributions,
because many of the dissolved particles rapidly (within a few seconds) proceed
to the reference sensor. To a smaller extent (and somewhat delayed) this
affects also the other sensor hidden in the trap. Then, the structure at larger
times contains our predicted effect among other heating and cooling
contributions (see above). However, the strongly significant drop between
some 50 through 70~min on the abscissa can certainly not be interpreted as
a process that warms up the reference sensor. We suggest that this decrease
quantifies the expected effect. Based on a fairly cautious interpretation
of the measurement statistics we observe a thermovoltage difference
around $\Delta U = 4.5$~$\mu$V. The accumulated thermal energy gradually decays
by cooling down to the temperature of the liquid right underneath
the particle trap. Note in Fig.~3 that all of our temperature variations
really are minor, and so interactions with the environment (such as
radiative heat loss or uptake) are insignificant contributions.

We cannot expect this type of recording (e.g., the plot in the upper panel)
at the end to level off at the same value as in the beginning,
since thermal conduction in water across the Petri dish
happens rather slowly. Equilibration over larger distances
is clearly beyond the plotted period of time. Solutions $T({\bf x},t)$ of
the thermal conduction equation for simplified model geometries confirm that
this is really so. On a significantly longer time scale (a few hours beyond
the plotted data) there is equilibration, and expectedly we measure a curve
that gradually approaches its inital level.

In addition, we observe the important fact that the time scale of the
achieved effect, i.e., of the structure centered at $t \approx 45$~min, agrees
well with the clearly visible penetration of the Brownian particles into the
trap. In view of the strongly retarded diffusion mechanism inside the trap
it appears almost evident that such measurements quite sensitively depend
on the geometrical details of the plug.

The bottom panel displays the very same type of measurement, except that
the plug of glass wool is some 0.3~cm longer. In the top curve the lower end
of the sensor coincides with the low end of the glass tube, and so the bottom
situation refers to a sensor that is 0.3~cm higher up. Our suggested mechanism
of heat accumulation due to frustrated Brownian motion does not even stand
such a small alteration, since then it takes too much time for the particles
to produce a measurable effect. Meanwhile cooling down has compensated the
heat accumulation, and so the outcome shows no time-dependence beyond the
initial trough. The relevant heating effect (that is absent now in the bottom
plot) proves reproducible, but actually it appears only in a rather small volume
of the parameter space that describes the totality of possible traps. Apart
from length and density of the plug, we are unable to name all the parameters
that matter, and different plugs with almost equal visible appearance may
well cause remarkably different results.

By virtue of the strong color, also for some modified plugs in the
spirit of the bottom example we aim at an observation of the progress
in propagation as the particles creep into the trap. As far as
possible, however, everything else (besides the plug length) is kept equal in
the two presented panels. In sum, we conclude that our suggested process of heat
transfer from outside into the trap works only marginally, thus raising some
difficulties to specify the the relevant geometrical criteria.

Moreover, if we compare the two panels we note that the strength of the minimum
at $t \approx 15$~min also quite sensitively depends on geometrical details. We
think the trough in the bottom curve is somewhat meager, but we have chosen this
plot because of its excellent long-term stability of the voltage measurement
conditions, which is not always the case. Apart from the mentioned geometry
aspects, we should also keep in mind that our plots in Fig.~3 represent
(among other contributions) small differences between energy quantities
of large magnitude. The ionic solution involves two very strong
thermal effects, namely dissociation and hydration that together amount to
quite a moderate solution heat.

Another purpose of the bottom panel is to guarantee that the relevant
structure in the upper curve is not induced by, e.g., heat production
due to oxidation of some contamination in the glass wool.
Otherwise, the bottom curve would exhibit this feature as well.
Similarly, we can rule out the possibility that our observed thermal effect
is caused by adsorption of Brownian particles at the glass fibers.
In the upper panel, one might also consider a possible retarded heat
flow out of the particle trap (that is now assumed to undergo insignificant
heating effects) into the cooler surrounding liquid. But, again, the bottom
panel does not support or admit this version to explain the upper measurement.

Furthermore, we think such plots as the bottom one strongly confirm our
imagination that the relevant thermal feature in the upper panel really
constitutes an entropy sink, and not just an entropy transfer within the
Petri dish (or even into the environment). The thermal compensation flow at
50 through 70~min produces entropy and restores the "standard" situation,
namely the one that throughout the recorded period of time has
never been left in the bottom panel. Altogether, in our view it is
justified to interpret the drop of some 4.5~$\mu$V in the upper curve to be
a measure for our expected effect of dissipated kinetic particle energy.

As we have indicated above, the inset in Fig.~3 presents an "absolute"
temperature measurement inside the particle trap. The reference sensor is now
placed in a second Petri dish filled with water kept at constant (room)
temperature throughout the measurement. Just as in the other two curves,
initially dilution enthalpy predominates. But now this concerns only the
sensor inside the trap, and so this causes a noticeable peak. This heating
process, however, cannot last too long: the liquid underneath the trap is also
subject to some retarded cooling due to consumption of solution heat (see above),
which gradually becomes more important along the time axis. The subsequent
feature (towards $t = 60$~min) again displays the effect under consideration.
We definitely do not pretend the small increase of temperature there to be
significant. However, our predicted mechanism of heat accumulation is an effect
relative to the liquid in the Petri dish where the trap is located, and not
primarily one with respect to an external reference. In short terms,
the inset in Fig.~3 disregards all the temperature effects in the immediate
neighborhood of the particle trap.

\subsection{Quantitative and practical implications}
Now we are ready to quantify how much the observed thermovoltage effect of roughly
4.5~$\mu$V means in terms of energy density. First, we calculate the number of
molecules in 1~g water, namely $3.34 \times 10^{22}$ of them. Suppose now, one
out of 1000 water molecules gets replaced by a Brownian particle that undergoes
a "full stop" in the trap. This yields $3.34 \times 10^{19}$ particles per gram
that contribute a dissipated kinetic energy of $(3/2)kT$ each. At room
temperature we end up with an appreciable energy density of 0.20~J/g. Since
4.186~J/g correspond to a temperature increase of 1~K in water, our 0.20~J/g
are able to warm up the water by 0.048~K, which amounts to some $U = 1.9$~$\mu$V
based on our type of thermocouple. However, we measured more than twice as much,
i.e., more than two Brownian particles within a 10 by 10 by 10 cube of water
molecules have completely lost their translatory kinetic energy they previously
possessed in a time average. Clearly, this example serves just as an illustration
for the size of our measured effect, in fact there will be many particles that
dissipate only a small fraction of their kinetic energy. Based on our careful
above assumption "two out of 1000" we arrive at an energy density of 0.4~J/g if
water is our host medium for the particles. However, diffusion-like processes
always are inconveniently slow, see abscissae in Fig.~3. In terms of power, these
0.4~J/g are produced approximately within an hour. This is by far not sufficient
to drive a car, but we think, e.g., of applications like hydrogen extraction
from sea water in a huge power plant.

Up to now, we suppressed the inevitable step where we have to empty the trap and
reinject the Brownian particles. This way, in a time average they again adopt
a translatory energy of $(3/2)kT$ each, and again some percentage of the
altogether present particles enter the trap and deposit part of their energy
there, and so on. We shall not underestimate this type of engineering task,
but we think in all probability this should be possible. A suitably
shaped and dimensioned siphon tube might even be adequate, but in our
present version of the experiment the water (plus particles) flow therein
would be inadmissibly small. In this particular setup we also must organize
a way to compensate for the loss of plain water in the trap, in order to
keep upright the particle density gradient. Once this situation is established,
the trap causes an oppositely directed temperature gradient that, within
certain limits, does not keep off the particles. No matter how we get
there, this "feedback mechanism" for the Brownian particles ideally
may happen in a slow and reversible process, and so, in principle, there
will be no extra supply of energy due.

One might be tempted to object that filling the trap happens spontaneously,
thus augments the overall system's entropy, and so it requires external
intervention to empty the trap in a continuous or cyclic manner. However, in
the theory section we already substantiated that heat transfer into the trap is
just a marginally permitted tiny deviation from thermodynamic equilibrium. This
resembles the diffusion process itself, since the particle density gradient
actually drives the dynamics of dilution, although this just insignificantly
affects the equipartition value of $(3/2)kT$. Thus we deal with two
spontaneously running processes, namely (i) the one that irreversibly
tends to even out the particle density gradients, and (ii) the
creation of a temperature gradient that lowers the overall
entropy. The limit of infinitely slow removal and reinjection
of particles raises no energy costs at all, since then there is no
density gradient equilibration anymore. In any case, a tiny fraction
of $(3/2)kT$ per particle suffices to overcome a density gradient.
Thus we may organize things in a way that the summary effect
of (i) and (ii) above decreases the entropy.

The continuous equilibration of the achieved temperature difference
goes along with reinjection of the Brownian particles, since this carries
slightly more kinetic energy into the trap compared to what the captured
particles keep in their nearby equilibrium state. No matter whether the
thermal compensation flow happens in an irreversible conduction process
(e.g., as in our presented single-step experiment) or whether we run a partly
reversible thermal engine, the overall amount of entropy within the closed
system will stay constant. Just the rate of particle feedback and the thermal
energy flow in such a steady state will depend on many parameters, above all
geometrical ones.

We also suppressed the actual physical mechanism under consideration, namely
how the Brownian motion gets modified such that the particles dissipate
part of their translatory energy. To consider Brownian particles
confined within some limited cavity is not really new, we point
to \cite{faucheux,condamin,lancon}, but these authors largely focused on
other aspects of this matter. The strongly different time scales
within and outside the particle trap (see above) clearly indicate that
the trap greatly influences a particle trajectory ${\bf x}(t)$. The
geometrical constraints reduce the probability for a particle to enter the trap
and to move on therein. At given particle density this corresponds to a lower
entropy, compared to the same volume of liquid outside the trap. In a sense,
this acts as a repelling hurdle, and sometimes this is said to be an
entropic barrier the particles have to surmount \cite{reguera}.

The trap involves also a so-called depletion force \cite{mao}, for a more
recent approach consult Ref.~\cite{herring}. This force again relates to entropy,
since it emerges from a geometrical exclusion of host molecules (water) at certain
locations. Depletion effects generally are discussed in a context of
particles larger than our hydrated ions, but such phenomena definitely
should be seen as part of the overall interaction with the trap. However, the
dissipation process inside the plug of glass wool remains to be an intuitive
picture, since we cannot trace a single particle's path in its geometrical
details. The trap seems to handicap the particles to an extent where they must
slightly depart from the thermodynamic (quasi-) equilibrium with the host
medium, but they hardly may aggregate to stable clusters.

A microscopically sound description of the interaction between
a single particle (a hydrated ion) and the trap surely is beyond
our present possibilities, since this then should be a precise ab initio
calculation without parameters that optionally may account
for the suggested effect. In the theory section we already addressed the
problem, and the circumstance that our Brownian particles are hydrated ions
renders the intended calculation particularly elaborate. Thus we cannot
present a trustworthy result in a context similar to Fig.~1 for the
frustrated Brownian motion inside the trap. This also prevents us from
an approach in the spirit of \cite{seifert}, recall our pertinent
discussion in the theory section. We would like to simulate the
particle motion in small cavities of variable shape as well as in the trap
as a whole, but it appears obvious that this cannot be carried out readily.
Besides the observable (and partly known) changes in the diffusion
properties, we also expect further important quantities to be modified.
Such characteristic quantities ultimately relate to the particle dynamics,
we think, e.g., of scaling properties and velocity distribution spectra.
Some of these peculiarities certainly are crucial to type and relative
incidence of interactions with the internal surfaces of the trap.

Conversely, on a merely qualitative basis it appears reasonable
to assume inelastic contributions in such interactions. At least
we expect these dissipative processes to happen among other, truly
elastic scattering or reflection events. Altogether, we are
positively convinced that our imagination of dissipated kinetic particle
energy stands a closer inspection based on various analytical methods,
theoretical as well as experimental ones, where the theoretical ones may
well be large-scale enterprises. Most importantly, the temperature effect
attended with the energy accumulation has been measured.

\section{Conclusions and outlook}
First, we would like to keep our above promise to discuss our specific
experimental arrangement in a context of accounting and possible
time-dependent displacements of entropy. In a closed system the entropy stays
constant as long as only reversible processes (such as the operation of
a Carnot engine) are going on therein. Our mechanism of spontaneous heat
transfer from outside to inside the particle trap, i.e., from cold to warm,
may easiest be regarded as a time reversed movie of an irreversible process.
Thus, it clearly represents a local decrease of entropy (although in terms of
energy percentage this is just a marginal effect on top of a thermodynamic
equilibrium situation). This, just by itself, is not unusual, e.g.,
isothermal compression as it may be part of a Carnot working cycle is an
archetypal "specimen" of such processes. Whenever we force a system
to adopt a state with lower probability, the important relation
$S = S_0 + k \ln P$ tells us that we have lowered the system's entropy $S$.
However, apart from small fluctuations around a mean value of $S$, such
processes always are attended with an entropy increase elsewhere in the
environment. But in our investigated process of warming up the particle trap
this is definitely not the case, since we deal with an entirely internal affair
that does not involve the environment. Apparently the Brownian particles
manage to "tap" the energy content of the liquid outside the trap. Later on,
when heat accumulation necessarily gets terminated in our single-step
experiment, we observe a thermal compensation flow that reverses the previous
energy transfer, see upper panel in Fig.~3.

In our view, this decrease in temperature of the (filled) particle
trap is a conclusive argument for our suggested mechanism,
and we do not see an alternate way to explain the experimental
facts on plausible grounds. Therefore, we conceived the strong
impression that this experiment indeed contradicts the second law
of thermodynamics. Further, the system as Fig.~2 represents it should be
considered as a closed one, i.e., there is no exchange of heat or
of matter with the environment. We think this is quite well realized
in our simple experiment: at our minute temperature changes there is
little heat loss or uptake by virtue of radiative, conductive, or convective
effects.

Once we have accepted the possibility of spontaneous heat accumulation
at the cost of another (cooler) place inside a (still closed) considered
system, we think of applications that are strictly prohibited within
traditional thermodynamics. But, admittedly, our presented mechanism
evolves rather slowly. Primarily we would like to "borrow" some
accumulated energy from the total energy content of the system. We simply may
"use it up" by means of all sorts of reversible and irreversible processes,
since the achieved temperature difference permits us to run thermal engines.
Finally, when all the borrowed energy, e.g., mechanical one, is dissipated
again and exists now only in thermal form, we provide a feedback to the place
where we got it from. This way the initial state gets restored, and we
are ready to enter the next cycle of energy acquisition and dissipation.
In principle, i.e., in an unrealistically strict implementation, this type
of energy economy is just self-sufficient.

A closed laboratory system, particularly a small one, may be advantageous to
investigate the aspects of principle in a well-controlled manner. However, there
is a considerable shortcoming that, in turn, is even beneficial to large-scale
experiments (i.e., sufficiently large that the system may be regarded as
an open one). Some of our energy demands are such that we cannot immediately
dissipate and "recycle" them. An obvious example is to build a house. The bricks
in the upper floors keep their potential energy almost forever, and so there is
always an amount of energy that subtracts from the overall thermal situation.
Within chemistry and particularly in biochemistry we certainly find more
striking examples. In all probability man's evolution is not able to
level off at a steady state, and so these considerations in fact matter.
At present, however, to store energy in any manifestation other than thermal
one may greatly help to unburden the earth climate.

Our setup sketched in Fig.~2 is completely left-right symmetric, except for
the presence of the particle trap. Thus, without the trap all the temperature
effects would just cancel. Apart from the inset, everything we observe
in Fig.~3 is uniquely due to the particle trap. Now we attempt to settle the
crucial question, namely: why does our specific experimental
arrangement contradict the second law of thermodynamics while other
observations apparently do not? We think the answer lies in a peculiarity
of the Brownian motion that generally has been investigated in a somewhat
one-sided manner. The driving force that ultimately lends the translatory
energy to the Brownian particles (we do not consider internal energy
contributions) and that finally moves them into the particle trap arises from
"accidentally" occurring deviations from the time-averaged situation in the
host medium. It is essential to recognize that the Brownian fluctuations
inherently constitute also a systematic dislocation according to Eq.~(4)
in the theory section. This mechanism results in the macroscopically observable
phenomenon of diffusion and, provided a density gradient, it continuously
accumulates particles in the trap. Such a particle transfer into the trap
operates at extremely low energy costs, since a particle's kinetic energy
just marginally depends on the density of surrounding other Brownian
particles. Thus, in our view, it appears feasible to organize
compensation for these costs by means of the dissipated fraction of the
initial kinetic particle energy.

By virtue of the equipartition value $(3/2)kT$, the amount of a particle's
translatory energy of motion is equally important, but this is not
primarily what we are after. Our experimental arrangement purposely
detects an effect that is driven by statistically occurring off-equilibrium
situations: we aim at Brownian particles inside the trap, although they have
started (and have picked up a tiny share of their kinetic energy) elsewhere.
The trap then must be able to distinguish between the different size of particles
and thus provide an obstacle to the Brownian motion, but this is merely
a geometrical affair. On an empirical basis, Fig.~3 exemplifies that a setup
with the ingredients of Fig.~2 indeed has the power to serve as a touchstone
for the rarely occurring possible second law violations.

Further, it may be noteworthy that our particle trap is not at all meant to be
a so-called Maxwell's demon, since a momentaneous thermal fluctuation that
kicks a Brownian particle into some direction happens by itself. Likewise, the
dissipation of kinetic energy inside the trap occurs spontaneously as well.
Actually it is the purpose of our approach to get by without an external
agent that separates the fast particles of an ensemble from the slow ones.
We already substantiated that our system is a closed one (i.e., there is some
insignificantly small heat exchange with the environment), and therefore
such a demon anyway would be part of the system under consideration.
We measure a resulting effect that originates from the above discussed
thermally induced short-term variations or asymmetries appearing in
a particle's local environment. After all, they cause an ongoing supply of
further (equally fast) particles to the trap.

The crucial prerequisite to accomplish a temperature difference is
a spontaneous mechanism that transfers some particles other than the ones
from the host medium from outside into the trap, and the Brownian motion gets
that done. Once a particle is in, the trap must be able to cause a slight
departure from equilibrium with the host medium, in order to receive a small
fraction of the kinetic particle energy in thermal form. In a sense, the result
of these two issues constitutes the "opposite" of an irreversible process
such as the thermal energy flow that equilibrates different temperatures.
Our approach ultimately acts as a "filter" that selects and stores momentaneous
deviations from a thermodynamic equilibrium, although this appears to be only
a marginally admitted effect among the overall phenomenology attended with
Brownian dynamics. That is why we think the finally achieved
spontaneous creation of a temperature gradient is not really that far
from what we may expect to happen on well-established thermodynamical
grounds. All in all, apart from fundamental and theoretical aspects
involved in this subject matter, we think the present contribution
offers an entirely new look at our future possibilities to make use of
energy resources.

\begin{acknowledgments}
The author thanks M. Erbudak, R. Monnier, J. Osterwalder,
and U. Straumann for constructive discussions.
\end{acknowledgments}

\end{document}